\documentclass[conference]{IEEEtran}
\IEEEoverridecommandlockouts
\usepackage{cite}
\usepackage{amsmath,amssymb,amsfonts}
\usepackage{algorithmic}
\usepackage{graphicx}
\usepackage{textcomp}
\usepackage{xcolor}
\usepackage{bbm}
\usepackage{tikz}
\usepackage{color,soul}

\newtheorem{theorem}{Theorem}
\newtheorem{lemma}{Lemma}
\newtheorem{proposition}{Proposition}
\newtheorem{definition}{Definition}

\newtheorem{remark}{Remark}

\def\BibTeX{{\rm B\kern-.05em{\sc i\kern-.025em b}\kern-.08em
    T\kern-.1667em\lower.7ex\hbox{E}\kern-.125emX}}
\begin{document}

\title{Game of Ages                         
\thanks{We acknowledge support of the Department of Atomic Energy, Government of India, under project no. 12-R\&D-TFR-5.01-0500.}
}

\author{\IEEEauthorblockN{Kumar Saurav}
\IEEEauthorblockA{\textit{School of Technology and Computer Science} \\
\textit{Tata Institute of Fundamental Research}\\
Mumbai, India. \\
kumar.saurav@tifr.res.in}
\and
\IEEEauthorblockN{Rahul Vaze}
\IEEEauthorblockA{\textit{School of Technology and Computer Science} \\
\textit{Tata Institute of Fundamental Research}\\
Mumbai, India. \\
rahul.vaze@gmail.com}
}

\def\onehalf{\frac{1}{2}}
\def\etal{et.\/ al.\/}
\newcommand{\bydef}{\triangleq}
\newcommand{\tr}{{\it{tr}}}
\def\SNR{{\textsf{SNR}}}
\def\bydef{:=}
\def\bba{{\mathbb{a}}}
\def\bbb{{\mathbb{b}}}
\def\bbc{{\mathbb{c}}}
\def\bbd{{\mathbb{d}}}
\def\bbee{{\mathbb{e}}}
\def\bbff{{\mathbb{f}}}
\def\bbg{{\mathbb{g}}}
\def\bbh{{\mathbb{h}}}
\def\bbi{{\mathbb{i}}}
\def\bbj{{\mathbb{j}}}
\def\bbk{{\mathbb{k}}}
\def\bbl{{\mathbb{l}}}
\def\bbm{{\mathbb{m}}}
\def\bbn{{\mathbb{n}}}
\def\bbo{{\mathbb{o}}}
\def\bbp{{\mathbb{p}}}
\def\bbq{{\mathbb{q}}}
\def\bbr{{\mathbb{r}}}
\def\bbs{{\mathbb{s}}}
\def\bbt{{\mathbb{t}}}
\def\bbu{{\mathbb{u}}}
\def\bbv{{\mathbb{v}}}
\def\bbw{{\mathbb{w}}}
\def\bbx{{\mathbb{x}}}
\def\bby{{\mathbb{y}}}
\def\bbz{{\mathbb{z}}}
\def\bb0{{\mathbb{0}}}

\def\bydef{:=}
\def\ba{{\mathbf{a}}}
\def\bb{{\mathbf{b}}}
\def\bc{{\mathbf{c}}}
\def\bd{{\mathbf{d}}}
\def\bee{{\mathbf{e}}}
\def\bff{{\mathbf{f}}}
\def\bg{{\mathbf{g}}}
\def\bh{{\mathbf{h}}}
\def\bi{{\mathbf{i}}}
\def\bj{{\mathbf{j}}}
\def\bk{{\mathbf{k}}}
\def\bl{{\mathbf{l}}}
\def\bm{{\mathbf{m}}}
\def\bn{{\mathbf{n}}}
\def\bo{{\mathbf{o}}}
\def\bp{{\mathbf{p}}}
\def\bq{{\mathbf{q}}}
\def\br{{\mathbf{r}}}
\def\bs{{\mathbf{s}}}
\def\bt{{\mathbf{t}}}
\def\bu{{\mathbf{u}}}
\def\bv{{\mathbf{v}}}
\def\bw{{\mathbf{w}}}
\def\bx{{\mathbf{x}}}
\def\by{{\mathbf{y}}}
\def\bz{{\mathbf{z}}}
\def\b0{{\mathbf{0}}}
\def\opt{\mathsf{OPT}}
\def\on{\mathsf{ON}}
\def\off{\mathsf{OFF}}
\def\bA{{\mathbf{A}}}
\def\bB{{\mathbf{B}}}
\def\bC{{\mathbf{C}}}
\def\bD{{\mathbf{D}}}
\def\bE{{\mathbf{E}}}
\def\bF{{\mathbf{F}}}
\def\bG{{\mathbf{G}}}
\def\bH{{\mathbf{H}}}
\def\bI{{\mathbf{I}}}
\def\bJ{{\mathbf{J}}}
\def\bK{{\mathbf{K}}}
\def\bL{{\mathbf{L}}}
\def\bM{{\mathbf{M}}}
\def\bN{{\mathbf{N}}}
\def\bO{{\mathbf{O}}}
\def\bP{{\mathbf{P}}}
\def\bQ{{\mathbf{Q}}}
\def\bR{{\mathbf{R}}}
\def\bS{{\mathbf{S}}}
\def\bT{{\mathbf{T}}}
\def\bU{{\mathbf{U}}}
\def\bV{{\mathbf{V}}}
\def\bW{{\mathbf{W}}}
\def\bX{{\mathbf{X}}}
\def\bY{{\mathbf{Y}}}
\def\bZ{{\mathbf{Z}}}
\def\b1{{\mathbf{1}}}

\def\bbA{{\mathbb{A}}}
\def\bbB{{\mathbb{B}}}
\def\bbC{{\mathbb{C}}}
\def\bbD{{\mathbb{D}}}
\def\bbE{{\mathbb{E}}}
\def\bbF{{\mathbb{F}}}
\def\bbG{{\mathbb{G}}}
\def\bbH{{\mathbb{H}}}
\def\bbI{{\mathbb{I}}}
\def\bbJ{{\mathbb{J}}}
\def\bbK{{\mathbb{K}}}
\def\bbL{{\mathbb{L}}}
\def\bbM{{\mathbb{M}}}
\def\bbN{{\mathbb{N}}}
\def\bbO{{\mathbb{O}}}
\def\bbP{{\mathbb{P}}}
\def\bbQ{{\mathbb{Q}}}
\def\bbR{{\mathbb{R}}}
\def\bbS{{\mathbb{S}}}
\def\bbT{{\mathbb{T}}}
\def\bbU{{\mathbb{U}}}
\def\bbV{{\mathbb{V}}}
\def\bbW{{\mathbb{W}}}
\def\bbX{{\mathbb{X}}}
\def\bbY{{\mathbb{Y}}}
\def\bbZ{{\mathbb{Z}}}

\def\cA{\mathcal{A}}
\def\cB{\mathcal{B}}
\def\cC{\mathcal{C}}
\def\cD{\mathcal{D}}
\def\cE{\mathcal{E}}
\def\cF{\mathcal{F}}
\def\cG{\mathcal{G}}
\def\cH{\mathcal{H}}
\def\cI{\mathcal{I}}
\def\cJ{\mathcal{J}}
\def\cK{\mathcal{K}}
\def\cL{\mathcal{L}}
\def\cM{\mathcal{M}}
\def\cN{\mathcal{N}}
\def\cO{\mathcal{O}}
\def\cP{\mathcal{P}}
\def\cQ{\mathcal{Q}}
\def\cR{\mathcal{R}}
\def\cS{\mathcal{S}}
\def\cT{\mathcal{T}}
\def\cU{\mathcal{U}}
\def\cV{\mathcal{V}}
\def\cW{\mathcal{W}}
\def\cX{\mathcal{X}}
\def\cY{\mathcal{Y}}
\def\cZ{\mathcal{Z}}

\def\sfA{\mathsf{A}}
\def\sfB{\mathsf{B}}
\def\sfC{\mathsf{C}}
\def\sfD{\mathsf{D}}
\def\sfE{\mathsf{E}}
\def\sfF{\mathsf{F}}
\def\sfG{\mathsf{G}}
\def\sfH{\mathsf{H}}
\def\sfI{\mathsf{I}}
\def\sfJ{\mathsf{J}}
\def\sfK{\mathsf{K}}
\def\sfL{\mathsf{L}}
\def\sfM{\mathsf{M}}
\def\sfN{\mathsf{N}}
\def\sfO{\mathsf{O}}
\def\sfP{\mathsf{P}}
\def\sfQ{\mathsf{Q}}
\def\sfR{\mathsf{R}}
\def\sfS{\mathsf{S}}
\def\sfT{\mathsf{T}}
\def\sfU{\mathsf{U}}
\def\sfV{\mathsf{V}}
\def\sfW{\mathsf{W}}
\def\sfX{\mathsf{X}}
\def\sfY{\mathsf{Y}}
\def\sfZ{\mathsf{Z}}

\def\bydef{:=}
\def\sfa{{\mathsf{a}}}
\def\sfb{{\mathsf{b}}}
\def\sfc{{\mathsf{c}}}
\def\sfd{{\mathsf{d}}}
\def\sfee{{\mathsf{e}}}
\def\sfff{{\mathsf{f}}}
\def\sfg{{\mathsf{g}}}
\def\sfh{{\mathsf{h}}}
\def\sfi{{\mathsf{i}}}
\def\sfj{{\mathsf{j}}}
\def\sfk{{\mathsf{k}}}
\def\sfl{{\mathsf{l}}}
\def\sfm{{\mathsf{m}}}
\def\sfn{{\mathsf{n}}}
\def\sfo{{\mathsf{o}}}
\def\sfp{{\mathsf{p}}}
\def\sfq{{\mathsf{q}}}
\def\sfr{{\mathsf{r}}}
\def\sfs{{\mathsf{s}}}
\def\sft{{\mathsf{t}}}
\def\sfu{{\mathsf{u}}}
\def\sfv{{\mathsf{v}}}
\def\sfw{{\mathsf{w}}}
\def\sfx{{\mathsf{x}}}
\def\sfy{{\mathsf{y}}}
\def\sfz{{\mathsf{z}}}
\def\sf0{{\mathsf{0}}}

\def\Nt{{N_t}}
\def\Nr{{N_r}}
\def\Ne{{N_e}}
\def\Ns{{N_s}}
\def\Es{{E_s}}
\def\No{{N_o}}
\def\sinc{\mathrm{sinc}}
\def\dmin{d^2_{\mathrm{min}}}
\def\vec{\mathrm{vec}~}
\def\kron{\otimes}
\def\Pe{{P_e}}
\newcommand{\expeq}{\stackrel{.}{=}}
\newcommand{\expg}{\stackrel{.}{\ge}}
\newcommand{\expl}{\stackrel{.}{\le}}
\def\SIR{{\mathsf{SIR}}}

\def\nn{\nonumber}

\maketitle

\begin{abstract}
We consider a distributed IoT network, where each node wants to minimize its age of information and there is a cost to make any transmission. A collision model is considered, where any transmission is successful from a node to a 
common monitor if no other node transmits in the same slot. There is no explicit communication/coordination between any two nodes. The selfish objective of each node is to minimize a function of its individual age of information and its transmission cost. Under this distributed competition model, the objective of this paper is to find a distributed transmission strategy for each node that converges to an equilibrium. The proposed transmission strategy only depends on the past observations seen by each node and does not require explicit information of the number of other nodes, or their strategies. 
A simple update strategy is shown to converge to an equilibrium, that is in fact a Nash equilibrium for a suitable utility function, that captures all the right tradeoffs for each node.  In addition, the price of anarchy for the utility function is shown to approach unity as the number of nodes grows large.
\end{abstract}


\section{Introduction} Consider the modern IoT paradigm in a 5G context, where there are large number of small IoT devices spread across a medium sized environment, e.g., a home, an office, an automobile or a factory floor. Each IoT device is monitoring certain inputs, and wants to communicate an update to a common monitor essentially as soon as possible. To model this scenario, a metric called the age of information (AoI) was introduced recently, that represents the freshness of information at the monitor/ receiver side, that has become a very popular object of theoretical interest in recent past \cite{kaul2012real, huang2015optimizing, sun2017update, yates2016age, najm2018content}. A nice review can be found in \cite{kosta2017age}.  Essentially, the age for any device at any time is the difference between the current time and the generation time of the last update.

Many variants of the AoI problem for a single node, e.g. depending on the scheduling discipline like FCFS \cite{kaul2012real}, or LCFS \cite{kaul2012status} and more importantly with multiple nodes has been considered in prior work, e.g., with multiple sources in \cite{yates2016age, huang2015optimizing, najm2018content, kadotascheduling, sun2018age, hsu2017scheduling, tripathi2017age}. 
With multiple sources, at each time slot, one bit of information can be sent from a set $S$ of sources to a monitor, e.g., $|S| = 1$ in \cite{kadotascheduling}, and the objective is to minimize the long-term weighted sum of the ages of all sources, subject to individual source throughput constraints. 

One common assumption between almost all prior work on AoI with multiple nodes is the centralized control over the transmission decisions by each node. For example, the policy in  \cite{kadotascheduling}, transmission decisions for each node are based on the current global age for each source. The centralized policies lead to a large overhead and delay, which could be limiting in a practical large scale IoT deployment, where devices are low-powered and delay sensitive, and a distributed or autonomous setup is preferred, where each IoT device can make its own decisions, given the transmission history.

This paper focusses on the distributed IoT paradigm, where each device has to make autonomous decisions with no communication between nodes. To keep the model simple, we consider slotted time, and assume that if two nodes transmit in the same slot, a collision occurs, and no update is received at the monitor. To keep the model practical, we assume that each node incurs a fixed cost for each transmission, thus ensuring that no node can transmit all the time. 
Under this distributed setting, the objective of each node is to minimize its own time-averaged age of information while incurring a reasonable average cost of transmission.


A typical approach to study such problems is to model it as a game, with a particular utility function, and then try to find a Nash equilibrium (NE) for it. There are multiple issues with such an approach: $(i)$ the choice of exact utility function is not obvious, and more importantly $(ii)$ gathering network information : e.g. the knowledge of number of other nodes may not be available in a distributed setting, possibly because of time varying nature, etc. The solution concept of NE is important in a distributed competitive setting, since it establishes that there is a fixed point or a stable strategy for each node, 
and ensures that the system can be driven to an {\it equilibrium}. 

In this paper, to eliminate the need for network information, we take an alternate approach to reach equilibrium via considering a local probabilistic transmit (learning) algorithm for each node, which decides the probability with which each node transmits its most recent packet in any slot. 
The learning algorithm for each node is local in the sense that it only depends on its own current time-averaged age, current average transmission cost, and past history of success/failures in slots in which it had transmitted a packet. 
With this learning algorithm, the objective is to show that it converges to a fixed point/equilibrium when followed by all nodes autonomously. 



%

In prior work, finding learning algorithms that achieve equilibrium has been considered for {\it congestion} games (that are also potential games) where the congestion costs are additive, and the multiplicative weights learning algorithm is known to converge to NE \cite{kleinberg2009multiplicative, krichene2015online}. For a more general setting, Friedman and Shenker \cite{friedman1997learning} showed that learning algorithms can achieve the NE in a two player zero-sum game, however, a similar result does not hold for a three player game as shown by Daskalakis et al. \cite{daskalakis2010learning}. For a brief survey, we refer the reader to the work of Shoham et al. \cite{shoham2007if}. 
For non-congestion games, learning algorithms achieving the NE has been briefly considered \cite{altman1998individual, kasbekar2010opportunistic, chen2013distributed}. 
Similar to our setup, there is also work \cite{marden1,marden2} in finding utility functions for which a given set of strategies are NE. Finding utility functions, however, for which the given set of strategies in addition have low price of anarchy is something that has remained intractable. 

The most related work on learning algorithms to achieve equilibrium for communication settings  is \cite{reverseMAC, thaker2017arrive, kavitha}. In particular, for analyzing exponential backoff  \cite{reverseMAC} and for arrival games \cite{thaker2017arrive}, existence of equilibrium via a learning algorithm is established. 
In \cite{kavitha}, an uplink throughput game is considered, where in a distributed setup each node is interested in maximizing its throughput via updating its transmission rate using a learning algorithm. Notably, \cite{kavitha} shows that it is not always possible to show the existence of an equilibrium or how to achieve it, and in principle, the learning algorithm based approach to achieve equilibrium is challenging.

In our model, we consider that each IoT device always has a packet to transmit following \cite{kadotascheduling, sun2018age, hsu2017scheduling}. Under the presence of multiple competing nodes, and the collision model, it is not clear when should each node transmit without any explicit communication between any two nodes, and when there is a cost for each transmission. Thus, a local probabilistic transmit (learning) algorithm is considered, where each node decides to transmit in each slot with probability that is determined by its own local knowledge of past successes and failures, current empirical age/cost etc, and the goal is to reach an equilibrium in this distributed setting. 
In particular, the proposed learning algorithm weights the current empirical average of age and cost inverse exponentially, which is intuitive, since larger the time-averaged age more aggressive should be the transmission probability and opposite for large average transmission cost. 
The fact that deterministic strategy cannot be an equilibrium strategy can be argued rather easily. 


The learning algorithm tries to find the right balance between transmitting too often that will lead to lot of collisions and large transmission cost, and transmitting too seldom which will increase the time-averaged age. Moreover, the learning algorithm does not need any knowledge of the network, for example, the number of other nodes in the network, transmission strategy of other nodes, etc. 

The {\bf main result} of this paper is to show that the proposed learning algorithm converges to a unique, non-trivial fixed point (equilibrium). We also explicitly characterize the fixed point, and show that it is in fact a NE for a game, where the corresponding {\it virtual} utility function captures the relevant tradeoffs of the problem, i.e., the utility function for each node is a function of its own transmission probability via the time-averaged age and average transmission cost, and is decreasing in the other nodes' transmission probabilities, etc. 

It is worth noting that the actual probabilistic learning algorithm makes no use of the knowledge of this {\it virtual} utility function that depends on network parameters such as the number of nodes in the network, and that is why we call it the {\it virtual} utility function. 
The virtual utility function is discovered only to characterize the fixed point of the learning algorithm. 
Moreover, we are also able to show that the price of anarchy of the virtual utility function approaches $1$ as the number of nodes grows large. This shows that even if nodes knew the virtual utility function and could collaborate, the optimal social utility would be close to the sum of the utilities obtained by the proposed learning algorithm.



The main technical ingredients of the paper are as follows. 
We first consider an expected version of the proposed learning algorithm, where all random variables are replaced by their expected values. We then find the underlying virtual utility function that the expected learning algorithm is trying to maximize. Corresponding to this utility function, we identify a multiplayer game $\cG$, and show that there is a unique NE for this game, and that is achieved by the best response strategy. 
To show the convergence of the proposed learning algorithm to a fixed point, we show that its updates converges to the best response actions for $\cG$. Thus, in two steps: $(i)$  proposed learning algorithm converges to the best response actions for $\cG$, and $(ii)$ best response actions for $\cG$ converges to the NE for the $\cG$, we show that the proposed learning algorithm converges to a fixed point characterized by the NE of $\cG$.
Note that this correspondence between the learning algorithm and the best response strategy that required network information is only made for analysis, and the learning algorithm does not need any network information.

We also present numerical results to validate our theoretical findings. In particular, we show that the proposed learning algorithm converges to an equilibrium quite fast, and happens for any choice of $N$, the number of nodes in the network. 
To show this effect, we perturb the system by increasing/decreasing $N$, and plot the resultant transmission probabilities. 
We plot the time-averaged age seen by any node in the network, which appears to grow exponentially with $N$ as expected, since there is no coordination in the network, and the success probability for node $i$ is $p_i\prod_{j\ne i, j=1}^N(1-p_j)$ if $p_i$ is the fixed point for each node $i$.
Even though we analytically only show that the price of anarchy approaches $1$ as the number of nodes become large, in simulations, we observe that it is very close to $1$ for all values of the number of nodes in the network.

\section{System Model}
\begin{figure}[htbp]
	\begin{center}
		\begin{tikzpicture}[thick,scale=0.9, every node/.style={scale=1}]
		\draw[->] (-0.25,0) to (7.5,0) node[below]{slot($i$)};
		\draw[->] (0.6,-0.25) to (0.6,2.5) node[below left]{$\Delta_\ell(i)$};
		\draw (0.6,0.25) to (2,1.65) to (2,0) to (4,2) to (4,0) to (5.75,1.75) to (5.75,0) to (6.25,0.5); 
		
		\draw[|<->|] (0.4,0.25) node[left]{$\Delta_{\ell0}$} -- (0.4,1.65) node[rectangle,inner sep=-1pt,midway,fill=white]{$i_1$};
		
		\draw[loosely dotted] (6.5,1) to (7.3,1); 
		
		\draw (2,-0.1) node[below]{$i_1$} to (2,0.1);
		\draw (4,-0.1) node[below]{$i_2$} to (4,0.1);
		\draw (5.75,-0.1) node[below]{$i_3$} to (5.75,0.1); 
		
		\end{tikzpicture}
		\caption{Sample plot of age (of node $\ell$) against slot.}
		\label{fig:age} 
	\end{center}
\end{figure}
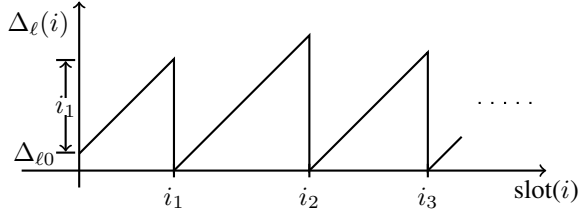

Consider a network with $N$ nodes and a single receiver/monitor. Time is discretized into equal-length slots. 
Following prior work \cite{ kadotascheduling, sun2018age, hsu2017scheduling, tripathi2017age}, we assume that a new data-packet (in short, packet) is generated in each slot at each node. 
If a node decides to transmit in any slot, it transmits the most recent packet, irrespective of success/failures of transmission in earlier slots.
Packet transmitted by a node in a slot is correctly decoded by the monitor if no other node transmits in the same slot. Otherwise, a collision occurs and all the simultaneously transmitted packets are lost. For realistic modelling, 
we assume that each transmission by node $\ell$ costs $c_{\ell}$ units, to capture transmit energy cost, etc. 
Also, in each slot $i$, the age of node $\ell$ is given by $i-\mu_{\ell}(i)$, where $\mu_{\ell}(i)$ is the last slot (relative to slot $i$) in which the packet of node $\ell$ was successfully received by the monitor. 
Fig. \ref{fig:age} shows a sample plot of age against slot. 
Here, $\Delta_\ell(i)$ denotes the age of node $\ell$ in slot $i$, while $\Delta_{\ell 0}$ is its initial age. Until the monitor receives a packet from node $\ell$, the age of node $\ell$ grows linearly with passage of slots, and it drops to 0 when the packet is received. 

Since nodes are distributed and there is no coordination/communication between them, a natural competition model emerges. Each node wants to transmit often to minimize its age, but in distinct slots, since otherwise there is collision in which all the nodes (colliding) accrue transmission cost but without any age reduction. Thus, each node wants to inherently selfishly maximize 
a {\it utility function} that depends on its time-averaged age and average transmission cost. The most appropriate form of utility function is debatable, and even if we consider a specific utility function, analytically showing that a NE exists and can be achieved may not be tractable. The basic idea behind seeking a NE is to show that there is a fixed point or a stable strategy for each node, and system can be made to work in an {\it equilibrium}. 

In this paper, we take an alternate approach to reach equilibrium via considering a local probabilistic learning algorithm (learning algorithm called hereafter) for each node, which decides the probability with which each node decides to transmit its most recent packet in any slot. 
The learning algorithm for each node is local in the sense that it only depends on its own current time-averaged age, current average transmission cost, and past history of success/failures in slots in which it had transmitted a packet. This approach completely eliminates the need for network parameter knowledge, such as the number of other nodes in the network, and their strategies, which would be needed if we were to follow the usual technique of considering an utility function for each node, and finding its NE.

With this learning algorithm, the objective is to show that it converges to a fixed point/equilibrium when followed by all nodes autonomously. 
A priori this appears to be a challenging task, however, we show in the next section, that it is possible to do so for the considered problem. In fact, we also characterize the fixed point that this learning algorithm achieves.

%

\section{The Learning Algorithm}

Let $m$ be a positive integer. Divide time-axis into frames by grouping $m$ consequtive slots together. Therefore, $t^{th}$ frame consists of slot $m(t-1)+1$ to slot $mt$. Henceforth, let $(t,i)$ refer to slot $m(t-1)+i$, i.e., $i^{th}$ slot in the $t^{th}$ frame. 
Further, let $C_{\ell}^{av}(t)$ be the average transmission cost of node $\ell$ in frame $t$, given by 
\begin{equation}\label{eq:Cav}
C_\ell^{av}(t)=\frac{c_{\ell}\sum_{i=1}^{m}T_\ell(t,i)}{m},
\end{equation}
where $c_{\ell}$ is the cost per transmission for node $\ell$, and $T_\ell(t,i)$ is a binary random variable which takes value 1 if node $\ell$ transmits in slot $(t,i)$, and 0 otherwise. Also, let $\Delta_\ell^{av}(t)$ be the time-averaged age of node $\ell$ in frame $t$ (assuming age at the start of the frame to be 0), given by  
\begin{equation}\label{eq:avAge}
\Delta_\ell^{av}(t)=\frac{\sum_{i=1}^{m}\Delta_\ell(t,i)}{m},
\end{equation} 
where $\Delta_{\ell}(t,i)=min\{i,\mu_{\ell}(t,i)\}$ is the age of node $\ell$ in slot $(t,i)$ where age is reset at the start of frame $t$ to be 0 (recall that $\mu_{\ell}(t,i)$ denotes the last slot (relative to slot $(t,i)$) in which the packet of node $\ell$ was successfully received by the monitor).

Consider the following learning algorithm for deciding the probability with which to transmit in any slot in a given frame: in each slot of a frame $t$, node $\ell$ transmits the packet with probability $p_{\ell}(t)$, where $p_{\ell}(1)$ is initialized with a random value from interval $(0,1)$, whereas at the end of each frame $t\ge 1$, $p_{\ell}(t+1)$ is given by 
\begin{align} \label{eq:algo}
p_\ell(t+1)=max\{&p_\ell^{min}, p_\ell(t)+\kappa(t)(e^{-\rho_{\ell 1} C_\ell^{av}(t)}\nonumber \\&-\frac{1}{(1+\Delta _\ell^{av}(t))e^{\rho_{\ell 2}}}-p_\ell(t))\},
\end{align} 
where, $\kappa(t)\in (0,1]$ is the learning rate, while   $p_{\ell}^{min}\in(0,1)$ and $\rho_{\ell 1},\rho_{\ell 2}>0$ are constants decided by node $\ell$ locally (i.e., independently of other nodes in the network).


The intution for \eqref{eq:algo} is as follows. If $C_\ell^{av}(t)$ is high, $p_\ell(t+1)$ should decrease, whereas, if $\Delta_\ell^{av}(t)$ is high, $p_\ell(t+1)$ should increase. To account for this trade-off, the learning component of algorithm \eqref{eq:algo} consists of two additive terms: $(i)$ $e^{-\rho_{\ell 1} C_\ell^{av}(t)}$ (a decreasing function in $C_\ell^{av}(t)$), and $(ii)$ $-1/((1+\Delta_\ell^{av}(t))e^{\rho_{\ell 2}})$ (an increasing function in $\Delta_\ell^{av}(t)$ due to negative sign). Because $C_\ell^{av}(t)\in[0,1]$ and $\Delta_\ell^{av}(t)\in[0,\infty)$, so the value of first term belongs to the interval $[e^{-c\rho_{\ell 1}},1]$ while the value of second term belongs to the interval $[-e^{-\rho_{\ell 2}},0)$. Therefore, $\rho_{\ell 1}$ and $\rho_{\ell 2}$ controls the relative weights of $C_\ell^{av}(t)$ and $\Delta_\ell^{av}(t)$  respectively, as well as the range of values the corresponding terms can take.

Note that by simple rearrangement of terms in \eqref{eq:algo}, we obtain $p_\ell(t+1)=max\{p_\ell^{min}, \kappa(t)(e^{-\rho_{\ell 1} C_\ell^{av}(t)}-\frac{1}{(1+\Delta _\ell^{av}(t))e^{\rho_{\ell 2}}})+(1-\kappa(t))p_\ell(t)\}$.  
	For each $t\ge 1$, $e^{-\rho_{\ell 1} C_\ell^{av}(t)}-1/((1+\Delta_\ell^{av}(t))e^{\rho_{\ell 2}})<1$, and because we initialize $p_{\ell}(1)<1$, therefore for $t\ge 1$ and $\kappa(t)\in(0,1]$,  we have $\kappa(t)(e^{-\rho_{\ell 1} C_\ell^{av}(t)}-1/((1+\Delta_\ell^{av}(t))e^{\rho_{\ell 2}}))+(1-\kappa(t))p_{\ell}(t)<1$ (because convex combination of two terms with value in interval $(0,1)$ also lies in the same interval). Also, $max\{\cdot\}$ function ensures that $p_\ell(t+1)\ge p_\ell^{min}>0$. Hence $\forall t$, $p_\ell(t+1)\in [p_\ell^{min},1)$. And  as $[p_\ell^{min},1)\subseteq[0,1]$, therefore, $p_\ell(t+1)$ is a valid  transmission probability.

\begin{remark}
	For analytical tractability, we update the transmission probability for frame $t+1$, i.e., $p_\ell(t+1)$ in \eqref{eq:algo}, using the $C_\ell^{av}(t)$ and $\Delta_\ell^{av}(t)$, which only accounts for the average transmission cost and time-averaged age of the previous (i.e., $t^{th}$) frame (instead of all the previous frames). 
	We show that choosing large enough frame length $m$, this simplification is sufficient. 
\end{remark}


Now, with the given description of the learning algorithm, the three main results of the paper can be summarized as follows.
\begin{theorem} \label{thm:algo}
	If all the nodes in the system obtain their transmission probability using the learning algorithm \eqref{eq:algo}, then their transmission probabilities converge to a unique fixed point almost surely.
\end{theorem}
Theorem \ref{thm:algo} establishes that the learning algorithm \eqref{eq:algo} can achieve an equilibrium. Next, we  characterize this equilibrium as follows.

\begin{theorem} \label{thm:NE-fixedPt}
	The unique fixed point of Theorem \ref{thm:algo} corresponds to the NE of the non-cooperative game $\mathcal{G}=\{N, p_\ell, U_\ell; \ell \in [N]\}$, where the $N$ nodes act as $N$ players, and each node $\ell$ chooses an action $p_\ell \in [p_\ell^{min}, 1]$ to maximize its own (virtual) utility $U_\ell$ given by 
	\begin{align}\label{eq:utilfn}
	U_\ell(p_\ell;\mathbf{P}_{-\ell}) &= 
	-\frac{e^{-\alpha_{\ell} p_\ell}}{\alpha_{\ell}}-\frac{p_\ell^2}{2}\left(1+\frac{1}{b_\ell}\right)+\frac{1+\alpha_{\ell}}{\alpha_{\ell}},
	\end{align}
	where $\alpha_{\ell}=c_{\ell}\rho_{\ell 1}$, $b_\ell=\prod_{k \ne \ell}(1-p_k)^{-1}e^{\rho_{\ell 2}}$, and $\mathbf{P}_{-\ell}$ denotes the transmission probability of all the nodes in the system except node $\ell$.
\end{theorem}
The utility function for each node \eqref{eq:utilfn} is relevant for the considered problem since it is a function of its own transmission probability via the time-averaged age and average transmission cost, and is decreasing in the other nodes' transmission probabilities through $b_\ell$, etc.

Even though NE guarantees equilibrium but its efficiency is quantified by price of anarchy ($PoA$) that counts the price for selfish behaviour. For $U_\ell$ defined in \eqref{eq:utilfn}, let $U_{sys} = \sum_{\ell}U_\ell(p_\ell;\mathbf{P}_{-\ell})$ be the sum of the utilities of all nodes. Then PoA is defined as $PoA=\frac{U_{sys}(\mathbf{P}_{OPT})}{U_{sys}(\mathbf{P}_{NE})}$, where $\mathbf{P}_{OPT}$ is the global optimal for $U_{sys}$ while $\mathbf{P}_{NE}$ is the NE point. In the next Theorem, we show that the PoA remains close to $1$ (as desirable) for the considered game with the virtual utility function \eqref{eq:utilfn}. 

\begin{theorem} \label{thm:PoA-unity}
	For the non-cooperative game $\mathcal{G}$ defined in Theorem \ref{thm:NE-fixedPt}, as $N$ becomes large, the price of anarchy ($PoA$) approaches unity. 
\end{theorem}

The rest of the theoretical part of the paper is dedicated in proving the above three theorems.
We first consider an expected version of the proposed learning algorithm \eqref{eq:algo}, where all random variables are replaced by their expected values. We then find the underlying virtual utility function that the expected learning algorithm is trying to maximize. Corresponding to this utility function, we identify a multiplayer game $\cG$, and show that there is a unique NE for this game, and that is achieved by the best response strategy. 
To show the convergence of the proposed learning algorithm to a fixed point, we show that its updates converge to the best response actions for $\cG$. Thus, in two steps: i)  proposed learning algorithm converges to the best response actions for $\cG$, and ii) best response actions for $\cG$ converges to the NE of $\cG$, we show that the proposed learning algorithm converges to a fixed point characterized by the NE of $\cG$.


Let $\mathbf{P}(t)$ denote the transmission probability vector of the $N$ nodes. Then we have the following Lemma (proof in Appendix \ref{appendix:proof-lemma-av2exp}):
\begin{lemma} \label{lemma:av2exp}
	For a given $\mathbf{P}(t)$, if $m$ is large, 
	\begin{enumerate}
		\item $\begin{aligned}
		C_\ell^{av}(t) \xrightarrow{\text{a.s.}} c_{\ell} \cdot p_\ell(t),
		\end{aligned}$
		\item $\begin{aligned}
		\Delta_\ell^{av}(t) \xrightarrow{\text{a.s.}} \frac{1-p_\ell(t)\prod_{k \ne l}(1-p_k(t))}{p_\ell(t)\prod_{k \ne l}(1-p_k(t))}
		\end{aligned}$.
	\end{enumerate} 	
\end{lemma}
%

Replacing $C_\ell^{av}(t)$ and $\Delta_\ell^{av}(t)$ in \eqref{eq:algo} by their converged values (assuming large $m$ and using Lemma \ref{lemma:av2exp}), we obtain the following expected form of the learning algorthm \eqref{eq:algo}:
\begin{align} \label{eq:ExpAlgo}
p_\ell(t+1)  
= max\{p_\ell^{min}, & \ \ p_\ell(t)+\kappa(t)(e^{-\alpha p_\ell(t) } \nonumber \\ &-\frac{1}{b_\ell/p_\ell(t)}-p_\ell(t)) \},
\end{align} 
where, $\alpha_{\ell}=c_{\ell}\rho_{\ell 1}$, and $b_\ell=\prod_{k \ne \ell}(1-p_k)^{-1}e^{\rho_{\ell 2}}$. To avoid overload of notation, we use the same notation $p$ for this expected update strategy \eqref{eq:ExpAlgo} as in \eqref{eq:algo}, and the distinction will be clear in the sequel.
Note that \eqref{eq:ExpAlgo} is just for analysis and it cannot be used in practice as $b_{\ell}$ is unknown. 
Now using \eqref{eq:ExpAlgo}, we extract the virtual utility function that \eqref{eq:ExpAlgo} is trying to maximize. 
\begin{theorem} \label{thm:utilfn}
	Let $\mathbf{P}_{-\ell}$ denote the transmission probability of all the nodes in the system except node $\ell$. Then for a given $\mathbf{P}_{-\ell}$, the learning algorithm \eqref{eq:ExpAlgo} maximizes the following virtual utility function (unique upto a constant):
	\begin{align}\label{eq:utilfndummy}
	U_\ell(p_\ell;\mathbf{P}_{-\ell}) &= 
	-\frac{e^{-\alpha_{\ell} p_\ell}}{\alpha_{\ell}}-\frac{p_\ell^2}{2}\left(1+\frac{1}{b_\ell}\right)+\frac{1+\alpha_{\ell}}{\alpha_{\ell}}.
	\end{align}
    which is continuous and strictly concave for $p_\ell$ in interval $(0,1]$, with a unique maximizer $p_\ell^{*}$ which lies in $[e^{-\alpha_{\ell}}/2,1)$. 
\end{theorem}
%


\subsection{Non-Cooperative Game Model}\label{sec:GameModel}
Using the virtual utility function \eqref{eq:utilfndummy}, we next define a game, where the strategy of each user is the probability with which to transmit in each slot in an autonomous way.
Let $\mathcal{G}=\{N, p_\ell, U_\ell; \ell \in [N]\}$ be a game, with $N$ nodes as players, and each node $\ell$ chooses an action $p_\ell \in [p_\ell^{min}, 1]$ to maximize its own utility $U_\ell$ given by \eqref{eq:utilfndummy}. The best response of a node $\ell$ is given by 
\begin{equation} \label{eq:bestResponse}
p_\ell^{br}=\underset{p_\ell\in [p_\ell^{min},1]}{\arg\max} \hspace{1ex} U_\ell(p_\ell;\mathbf{P}_{-\ell}),
\end{equation}
and under best response strategy, at the end of each frame $t$, $p_{\ell}(t+1)=p_{\ell}^{br}$. Further, a \textit{Nash Equilibrium (NE)} is said to exist for $\mathcal{G}$ if there exists a transmission probability vector $\mathbf{P}^{NE}$, such that for each node $\ell$, $p_{\ell}^{NE}$ is best response for node $\ell$ given $\mathbf{P}_{-\ell}^{NE}$. 
Note that the set $\{p_\ell| p_\ell^{min} \leq p_\ell \leq 1\}$ is non-empty, compact and convex in $\mathbb{R}$. Additionally from Theorem \ref{thm:utilfn}, the utility function \eqref{eq:utilfndummy} is continuous and strictly concave (strict concavity implies quasi-concavity as well) for $p_\ell(t)\in[p_\ell^{min},1]$. Hence using Proposition \ref{prop:NEexist}, we conclude that NE exists for $\mathcal{G}$.
\begin{proposition}\label{prop:NEexist}
	\textnormal{[Proposition 20.3 in \cite{osborne1994course}]}
	The non-cooperative game $\mathcal{G}=\{N, p_\ell, U_\ell; \ell \in [N]\}$ has a Nash Equilibrium if for all $\ell\in[N]$,
	\begin{enumerate}
		\item the set of actions $\{p_\ell\}$ of player $\ell$ is a non-empty compact convex subset of a Euclidean space, and
		\item the utility function $U_\ell$ is continuous and quasi-concave on the set of actions $\{p_\ell\}$.
	\end{enumerate}
\end{proposition} 

Next, we show that the best response strategy \eqref{eq:bestResponse} for $\mathcal{G}$ converges to the unique NE.
\subsection{Convergence of Best Response Strategy}

\begin{theorem}\label{thm:BRS2NE}
	For the non-cooperative game $\mathcal{G}$, if for each node $\ell$, 
	\begin{align} \label{eq:BRS2NE}
	\frac{(N-1)(1-p_{global}^{min})^{(N-2)}}{e^{\rho_{\ell 2}}(\alpha_{\ell}+1)}< 1
	\end{align} 
	(where $p_{global}^{min}\le\underset{j}{min}\{p_j^{min}\}$), then the best response strategy converges to the unique NE. 
\end{theorem}

Theorem \ref{thm:BRS2NE} has been proved in Appendix \ref{appendix:proof-thm-BRS2NE}. Note that \eqref{eq:BRS2NE} depends on the value of $N$ and $p_{global}^{min}$. We have assumed that value of  $N$ is unknown to nodes. Next, we show that if each node chooses its parameters depending on a predetermined value of  $p_{global}^{min}$ independent of the value of $N$,  \eqref{eq:BRS2NE} can be made to satisfy for all values of $N$.
The result is summarized in the following Lemma (detailed proof in Appendix \ref{appendix:proofconvBRS2NE}).

\begin{lemma}\label{lemma:convBRS2NE}
	Let $p_{global}^{min}\in(0,0.5)$ be the lower bound on $p_{\ell}^{min}$ for each node $\ell$, known as part of the learning algorithm \eqref{eq:algo}. If each node $\ell$ assigns $p_{\ell}^{min}=p_{global}^{min}$, $\rho_{\ell 1}\le -\frac{1}{c_{\ell}}\ln(2p_{global}^{min})$, and  
	\begin{align}
	\rho_{\ell 2}> max \left\{0, \ln\left(\frac{(n^*-1)(1-p_{global}^{min})^{(n^*-2)}}{\alpha_{\ell}+1}\right)\right\},
	\end{align}
	where, $\begin{aligned}
	n^*=1-\frac{1}{\ln(1-p_{global}^{min})}, 
	\end{aligned}$ then \eqref{eq:BRS2NE} is always satisfied.
\end{lemma}

\begin{remark}
	As per Lemma \ref{lemma:convBRS2NE}, $\rho_{\ell 1}\le-\frac{1}{c_\ell}\ln(2p_{global}^{min})$.
	Suppose that for each node $\ell$, $\rho_{\ell 1}=-\frac{1}{c_\ell}\ln(2p_{global}^{min})$, then $\alpha_\ell=c_\ell\rho_{\ell 1}= -\ln(2p_{global}^{min})$, which is a global constant (independent of $c_\ell$). 
	Now from \eqref{eq:ExpAlgo}, note that for each node $\ell$, the trajectory of its transmission probability is determined by $\alpha_{\ell}$, $\rho_{\ell 2}$ and $p_\ell^{min}$. When $\alpha_{\ell}$ is independent of $c_\ell$, then as per Lemma \ref{lemma:convBRS2NE}, $\rho_{\ell 2}$ and $p_\ell^{min}$ are also independent of $c_\ell$. Therefore, if for each node $\ell$, $\rho_{\ell 1}=-\frac{1}{c_\ell}\ln(2p_{global}^{min})$, then the trajectory of transmission probability 
	is independent of $c_\ell$.
\end{remark}

In summary, we conclude that each node can independently choose the parameters such that the condition \eqref{eq:BRS2NE} for convergence of the best response strategy is satisfied for all values of $N$. 
  
To finally prove Theorem \ref{thm:algo}, we next show that the learning algorithm \eqref{eq:algo} converges to the best response strategy \eqref{eq:bestResponse} which converges to the NE of $\cal G$ as shown in Theorem \ref{thm:BRS2NE}.   
\subsection{Convergence of Learning Algorithm \eqref{eq:algo} to the Best Response Strategy \eqref{eq:bestResponse}} \label{sec:LA2BRS}

\begin{definition}\label{def:subgrad}
	For the real-valued concave function $U_\ell:(0,1]\rightarrow \mathbb{R}$, $\phi$ is said to be its subgradient at point $p_*\in(0,1]$, if for every other point $p_0\in(0,1]$, we have $U_\ell(p_0)-U_\ell(p_*)\le\phi\cdot (p_0-p_*)$. Further, a function $v(t)$ (where $t$ is time) is called the stochastic subgradient of $U_\ell(p_\ell;\mathbf{P}_{-\ell})$ at point $p_\ell(t)$, if for a given value of random variables $(p_\ell(0),p_\ell(1),...,p_\ell(t))$, $\mathbb{E}\{v(t)|p_\ell(0),p_\ell(1),...,p_\ell(t);\mathbf{P}_{-\ell}\}$ is a subgradient of $U_\ell(p_\ell;\mathbf{P}_{-\ell})$ at $p_\ell(t)$.
\end{definition}

Consider the function $v_\ell(t)$ defined as below:
\begin{align} \label{eq:subgrad} 
v_\ell(t)&=e^{-\rho_{\ell 1} C_\ell^{av}(t)}-\frac{1}{(1+\Delta _\ell^{av}(t))e^{\rho_{\ell 2}}}-p_\ell(t).
\end{align}

Taking conditional expectation on both sides of \eqref{eq:subgrad}, we get  
\begin{align}\label{eq:subgrad2slope}
\mathbb{E}\{v_\ell(t)|p_\ell(0),&p_\ell(1),...,p_\ell(t);\mathbf{P}_{-\ell}(t)\} \nonumber \\
&\stackrel{(a)}{=}e^{-\alpha_{\ell} p_\ell(t) }-\frac{1}{b_\ell/p_\ell(t)}-p_\ell(t), \nonumber\\
&\stackrel{(b)}{=} \frac{\partial U_\ell(p_\ell(t);\mathbf{P}_{-\ell}(t))}{\partial p_\ell(t)},
\end{align}
where (a) is obtained for large $m$ using Lemma \ref{lemma:av2exp} and using $\alpha_{\ell}$ and $b_{\ell}$ to denote $c_{\ell}\rho_{\ell 1}$ and $\prod_{k \ne \ell}(1-p_k)^{-1}e^{\rho_{\ell 2}}$ respectively, while (b) is obtained using \eqref{eq:utilfndummy}. Further due to Theorem \ref{thm:utilfn}, we know that $U_l$ is a concave function in $p_\ell$ (for fixed $\mathbf{P}_{-\ell}$). Therefore for $p_0,p_*\in(0,1]$ and $p_0\ne p_*$,
\begin{align} \label{eq:diff-subgrad}
U_\ell(p_0;\mathbf{P}_{-\ell})-U_\ell(p_*;\mathbf{P}_{-\ell})\le\frac{\partial U_\ell(p_\ell;\mathbf{P}_{-\ell})}{\partial p_\ell}\Big|_{p_\ell=p_*}\cdot (p_0-p_*).
\end{align} 
Hence, using \eqref{eq:subgrad2slope} and \eqref{eq:diff-subgrad} along with Definition \ref{def:subgrad}, we conclude that $v_\ell(t)$ is a stochastic subgradient of $U_\ell(p_\ell(t);\mathbf{P}_{-\ell}(t))$.
Now using \eqref{eq:subgrad}, we can write the learning algorithm \eqref{eq:algo} as
\begin{equation}\label{eq:subgradAlgo}
p_\ell(t+1)=max\{p_\ell^{min}, p_\ell(t)+\kappa(t)v_\ell(t)\},
\end{equation}
which suggests that the learning algorithm can be interpreted as a stochastic subgradient algorithm \cite{boyd2008stochastic} which maximizes the virtual utility function given by \eqref{eq:utilfndummy}. Using this interpretation of the learning algorithm we obtain Theorem \ref{thm:algo2br}, with detailed proof in Appendix \ref{appendix:proof-thm-algo2br}.

\begin{theorem} \label{thm:algo2br}
	If the learning rate $\kappa(\cdot)$ is chosen such that $ \forall t$, $\kappa(t)> 0$, $\sum_{t=1}^{\infty}\kappa(t)=\infty$, and $\sum_{t=1}^{\infty}\kappa^2(t)<\infty$, then the learning algorithm \eqref{eq:algo} converges to the best response strategy \eqref{eq:bestResponse} almost surely. 
\end{theorem}

Theorem \ref{thm:algo2br} suggests that for properly chosen learning rates (for example, $\kappa(t)=1/t,\forall t\ge 1$), the learning algorithm converges to the best response strategy almost surely, and if \eqref{eq:BRS2NE} is satisfied, then according to Theorem \ref{thm:BRS2NE}, the best response strategy further converges to a unique NE. Thus, combining Theorem  \ref{thm:BRS2NE} and Theorem \ref{thm:algo2br}, we conclude that if all the nodes update their transmission probabilities by following the learning algorithm \eqref{eq:algo} with an appropriate learning rate and satisfying \eqref{eq:BRS2NE}, then their transmission probabilities converge to a unique NE (a fixed point) almost surely, thereby proving Theorem \ref{thm:algo}. Thus, completing the proof of Theorem \ref{thm:algo} and Theorem \ref{thm:NE-fixedPt} simultaneously. Proof of Theorem \ref{thm:PoA-unity} can be found in Appendix \ref{appendix:proof-PoA-unity}.

\section{Numerical Results}

\begin{figure}[htbp]
	\centerline{\includegraphics[width=0.8\linewidth]{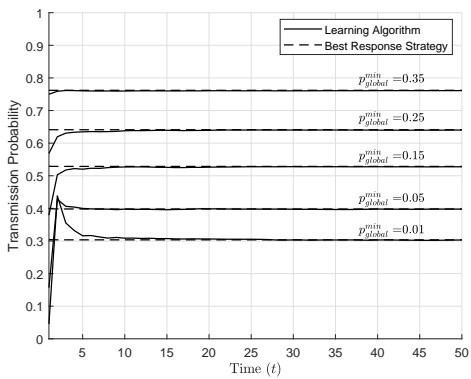}}
	\caption{Variation of transmission probability of a node with time.}
	\label{fig:Prob-Time}
\end{figure}

We analyzed the convergence properties of the learning algorithm \eqref{eq:algo} by simulating a scenario with 10 nodes and $\rho_{\ell 1}$, $\rho_{\ell 2}$, and $p_{\ell}^{min}$ chosen as per Lemma \ref{lemma:convBRS2NE} for different values of $p_{global}^{min}$. Also, $\kappa(t)=1/t,\forall t\ge 1$. As shown in Fig. \ref{fig:Prob-Time}, transmission probability obtained using the learning algorithm converges to the best response strategy very quickly.

\begin{figure}[htbp]
	\centerline{\includegraphics[width=0.8\linewidth]{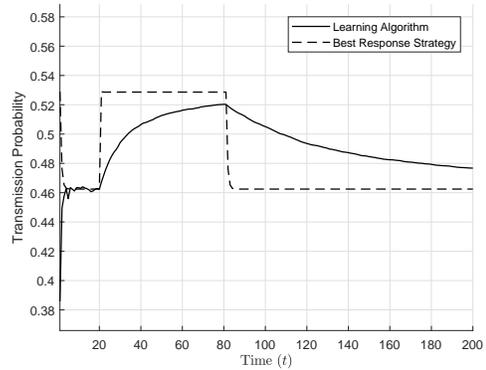}}
	\caption{Convergence of the learning algorithm to the best response strategy when number of nodes vary with time.}
	\label{fig:Prob-Time-disturb}
\end{figure}

To check the robustness of the learning algorithm under dynamic conditions, we performed a second simulation with 3 nodes at $t=0$ and 7 new nodes joining the system at $t=20$ and leaving it again at $t=80$. As shown in Fig. \ref{fig:Prob-Time-disturb}, irrespective of the disturbance, the learning algorithm converges to the best response strategy. But note that the learning rate $\kappa(t)$ decreases with $t$. Hence, if the system is disturbed at large $t$, then the convergence is slow. However, this issue can be resolved by reinitializing $t$ whenever it becomes very large.  

\begin{figure}[htbp]
	\centerline{\includegraphics[width=0.8\linewidth]{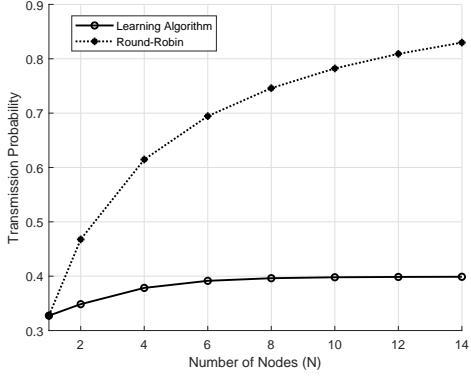}}
	\caption{Variation of transmission probability of a node with number of nodes.}
	\label{fig:Prob-N}
\end{figure}

To understand the effect of number of nodes $N$ on the fixed point of the learning algorithm \eqref{eq:algo}, Fig. \ref{fig:Prob-N} plots the transmission probability (converged) obtained using the learning algorithm \eqref{eq:algo} for different values of $N$ (for the simulation, we used $p_{global}^{min}=0.05$ and for each node $\ell$, $c_{\ell}=1$, while $p_{\ell}^{min}$, $\rho_{\ell 1}$ and $\rho_{\ell 2}$ were chosen as per Lemma \ref{lemma:convBRS2NE}). 
For comparative study, Fig. \ref{fig:Prob-N} also plots the transmission probability for the round-robin (RR) scheme, in which, each node is assigned a slot in round-robin fashion to avoid collision. With RR, the nodes may transmit their packets only in their respective alloted slots with transmission probability obtained using \eqref{eq:algo}. However, note that the RR is only of theoretical interest because in practice, there is no mechanism for slot allotment (as neither the nodes can communicate with each other, nor there is a centralized controller to do so).

\begin{remark}
When number of nodes is small, the interval between consequtive alloted slots of each node in RR is also small. Therefore, depending on the transmission cost of a node, it may not be optimal for the node to transmit packet in every alloted slot. Hence, to account for this fact, we consider that in RR, a node transmits in the alloted slot with probability obtained using \eqref{eq:algo}. Further, due to the specific choice of \eqref{eq:algo} for obtaining transmission probability under RR, the comparision of corresponding plots for the learning algorithm and RR provides nice insight regarding the impact of collision on the learning algorithm \eqref{eq:algo}. 
\end{remark}

For the learning algorithm \eqref{eq:algo}, as $N$ increases, there are two phenomena which simultaneously influence the transmission probability: $(i)$ For large $N$, frequency of packet collision is high. Therefore, average transmission cost increases with $N$, thereby decreasing the transmission probability. $(ii)$ With more collisions happening (and fewer packets getting received by the monitor) due to large $N$, time-averaged age becomes high, hence increasing the transmission probability.  
As shown in Fig. \ref{fig:Prob-N}, for small $N$, phenomenon $(ii)$ dominates, thereby increasing the transmission probability. However, when $N$ is large, the two phenomena balances each other, and hence, the transmission probability gets saturated. 

For round-robin scheme, as $N$ increases, interval between successive alloted slots of each node becomes large. Therefore for a fixed transmission probability, average transmission cost decreases, whereas time-averaged age increases: both leading to increase in transmission probability. Therefore, the transmission probability under RR increases very rapidly with $N$ (in comparision to the learning algorithm). 

\begin{figure}[htbp]
	\centerline{\includegraphics[width=0.8\linewidth]{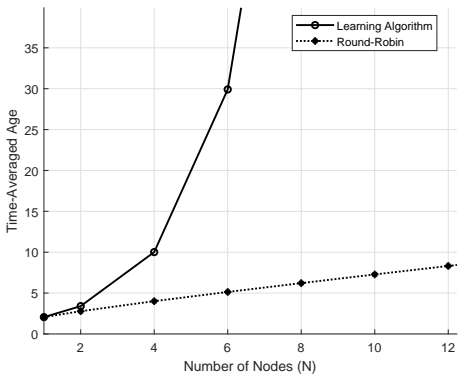}}
	\caption{Variation of time-averaged age with number of nodes.}
	\label{fig:zoomedAvAge-N}
\end{figure}

Additionally, we analysed the variation in time-averaged age with increase in $N$ for the learning algorithm as well as round-robin scheme. As shown in Fig. \ref{fig:zoomedAvAge-N}, time-averaged age for the learning algorithm increases very rapidly with $N$ in comparision to the round-robin scheme. 
If a packet from node $\ell$ is successfully received by the monitor once in every $T$ slots, then using \eqref{eq:avAge} and assuming $m/T$ ($m$ is the number of slots in each frame) to be an integer,  we get the time-averaged age to be 
\begin{align}
\Delta_{\ell}^{av}(t)=\frac{\sum_{j=1}^{m/T}\sum_{i=1}^{T}\Delta_{\ell}(t,i)}{\sum_{j=1}^{m/T}T}=\frac{T^2/2}{T}=\frac{T}{2}.
\end{align}
In round-robin scheme, a packet is successfully received every $N\mathbb{E}[s_{\ell a}]$ slots, where $s_{\ell a}$ is the number of alloted slots per transmission for node $\ell$. As shown in Fig. \ref{fig:Prob-N}, transmission probability increases with $N$, and hence, $\mathbb{E}[s_{\ell a}]$ decreases (approaches 1) as $N$ increases. Therefore, increase in time-averaged age for the learning algorithm \eqref{eq:algo} is $N\mathbb{E}[s_{\ell a}]/2$, which converges to $N/2$ when $N$ is large. Fig. \ref{fig:zoomedAvAge-N} shows a similar trend as can be verified using the transmission probability values from Fig. \ref{fig:Prob-N}. 

Now for the learning algorithm \eqref{eq:algo}, probability that a packet of node $\ell$ is received by the monitor is $p_{\ell}\prod_{k \ne \ell}(1-p_k)$. 
Let the transmission probability of each node to be equal (say, $p$). Therefore, the probability that a packet of node $\ell$ is received by the monitor becomes $p(1-p)^{N-1}$, and 
hence, the expected number of slots required for each successful reception of packet by the monitor is $O((1-p)^{-N})$. So when $N$ increases, the time-averaged age for the learning algorithm grows exponentially.

Finally, we also computed the price of anarchy ($PoA$) for the utility function of each node being \eqref{eq:utilfn}. For any combination of transmission probability of nodes given by $\mathbf{P}$, the overall utility of the system is given by $U_{sys}(\mathbf{P})=\sum_\ell U_{\ell}(p_{\ell};\mathbf{P}_{-\ell})$, where $U_{\ell}(p_{\ell};\mathbf{P}_{-\ell})$ is the utility of node $\ell$. Therefore, $PoA$ of the learning algorithm is 
\begin{align} \label{eq:def-PoA}
PoA=\frac{U_{sys}(\mathbf{P}_{OPT})}{U_{sys}(\mathbf{P}_{LA})},
\end{align}
where, $\mathbf{P}_{OPT}$ is the optimal transmission probability vector which maximizes $U_{sys}(\cdot)$, while $\mathbf{P}_{LA}$ is the vector of (converged) transmission probabilities obtained using the learning algorithm \eqref{eq:algo}. 
Note that $PoA\ge 1$, and a value close to 1 indicates that the algorithm is close to optimal.

\begin{figure}[htbp]
	\centerline{\includegraphics[width=0.8\linewidth]{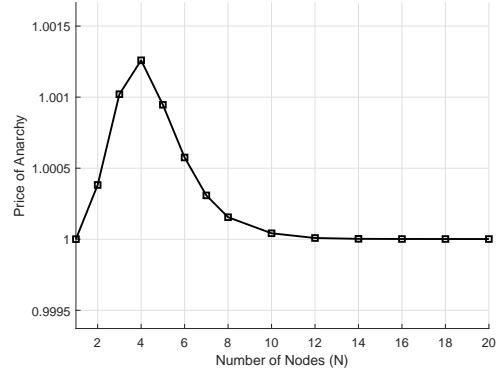}}
	\caption{Variation in Price of Anarchy $PoA$ of the Learning Algorithm with number of nodes.}
	\label{fig:PoA}
\end{figure}

Figure \ref{fig:PoA} plots the $PoA$ of learning algorithm \eqref{eq:algo} for different values of $N$. Initially when $N$ increases, $PoA$ increases as well. However, for large $N$, $PoA$ converges back to unity as per Theorem \ref{thm:PoA-unity}. Detailed explanation of the phenomenon is discussed in Appendix \ref{appendix:proof-PoA-unity}.

\section{Conclusion}
In this paper, we have presented a new direction in achieving equilibrium in a distributed IoT setting, 
where each node is interested in minimizing its age of information when there is a cost for each transmission. Typically, for distributed models, one identifies an utility function for each node and tries to establish a NE for it. However, such an approach requires the network knowledge, e.g., the number of nodes in the network, and their strategies, which may not be available in a distributed network. We instead propose a simple local update (learning) strategy for each node that determines the probability with which to transmit in each slot, that depends on the current empirical average of age and cost. This strategy for appropriate choice of parameters is shown to achieve an equilibrium that is also identified by a NE for a suitable virtual game. To further quantify the efficiency of this learning strategy, it is shown that the price of anarchy of the virtual game approaches unity when the number of nodes in the network is large enough.


\bibliographystyle{IEEEtran}
\bibliography{reflist,refs}
\appendices

\section{Proof of Lemma \ref{lemma:av2exp}} \label{appendix:proof-lemma-av2exp}

From \eqref{eq:Cav}, $C_\ell^{av}(t)=\frac{c_{\ell}\sum_{i=1}^{m}T_\ell(t,i)}{m}$, where $T_\ell(t,i)$ has Bernoulli distribution (takes value 1 with probability $p_\ell(t)$, and 0 otherwise). Since $T_\ell(t,1),T_\ell(t,2),...,T_\ell(t,m)$ are independent and identically distributed, therefore when $m$ is large, 
we get relation (1) using strong law of large numbers.

Now, note that 
$\Delta_\ell(t,i)=(\Delta_\ell(t,i-1)+1)\mathbbm{1}_{\{F_\ell(t,i)\}}$,
where $F_\ell(t,i)$ is the event that a packet transmitted by node $\ell$ is \emph{not} received by the monitor in slot $(t,i)$. Hence, 
\begin{align}\label{eq:expAge}
\mathbb{E}\{\Delta_\ell(t,m)|\mathbf{P}(t)\} = &[\mathbb{E}\{\Delta_\ell(t,m-1)|\mathbf{P}(t)\}+1] \nonumber \\ &[1-p_\ell(t)\prod_{k \ne \ell}(1-p_k(t))].
\end{align} 

We also have following Lemma  (proved in Appendix \ref{appendix:proofMarkovChain}):
\begin{lemma}\label{lemma:markov}
	For a fixed $t$, the sequence $\Delta_\ell(t,1), \Delta_\ell(t,2),...$ is an ergodic uniform Markov chain.
\end{lemma}

Using Lemma \ref{lemma:markov}, when $m$ is large, $\mathbb{E}\{\Delta_\ell(t,m)|\mathbf{P}(t)\}=\mathbb{E}\{\Delta_\ell(t,m-1)|\mathbf{P}(t)\}$, and additionally using the extension of Birkhoff ergodic theorem discussed in \cite{sandric2017note}, $\Delta_\ell^{av}(t)\xrightarrow{\text{a.s.}}\mathbb{E}\{\Delta_\ell(t,m)|\mathbf{P}(t)\}$. Therefore, plugging these results in \eqref{eq:expAge}, we get relation (2) of Lemma \ref{lemma:av2exp}. 

\section{Proof of Lemma \ref{lemma:markov}} \label{appendix:proofMarkovChain}
Within a frame $t$, $\mathbf{P}(t)$ is fixed, and hence,
\begin{align}
\mathbb{P}[\Delta_\ell(t,i+1)=x|\Delta_\ell(t,i)=y]=
\begin{cases}
1-\nu; & x=y+1\\ 
\nu; & x = 0
\end{cases} \nonumber
\end{align}
where $\nu=p_\ell(t)\prod_{k \ne \ell}(1-p_k(t))$ is the probability that the packet transmitted by node $\ell$ in a slot of frame $t$ is successfully received by the monitor. Therefore for a given state $\Delta_\ell(t,i)=y$, $\Delta_\ell(t,i+1)$ is written independently of $\Delta_\ell(t,j), \forall j<i$, and the transition probability is independent of $i$. Therefore, the sequence $\Delta_\ell(t,1), \Delta_\ell(t,2),...$ is a uniform Markov chain. Further, note that $\mathbb{P}[\Delta_\ell(t,i+x+1)=x|\Delta_\ell(t,i)=x]=\nu(1-\nu)^x>0$, as well as $\mathbb{P}[\Delta_\ell(t,i+x+2)=x|\Delta_\ell(t,i)=x]=\nu^2(1-\nu)^x>0$. Therefore, the Markov chain is also aperiodic (i.e., period$=1$). Hence, to prove that the Markov chain is a ergodic uniform Markov chain, it is sufficient to show that it positive recurrent.

Also, from any state $y$, any other state $x$ can be reached in finite number of steps (slots) with positive probability, given by $(1-\nu)^{x-y}$ if $x>y$, and $\nu(1-\nu)^x$ if $x\le y$.
So, the Markov chain is a single communicating class. Hence to show that it is positive recurrent, it is sufficient to show that any particular state is positive recurrent \cite{kumar2012discrete}.

Let $f_{00}^{(j)}$ denote the probability for returning to state $0$ in $j^{th}$ step. Then $\forall j\ge 1$, we have $f_{00}^{(j)}=\nu>0$. Hence,
\begin{equation}\label{eq:chainRecurrent}
\lim_{m\to \infty}\sum_{j=1}^{m}f_{00}^{(j)}=\infty, \hspace{1ex}\text{and} \hspace{1ex}\lim_{m\to\infty}\frac{1}{m}\sum_{j=1}^{m}f_{00}^{(j)}=\nu>0.  
\end{equation}

Therefore using Theorem \ref{thm:DTMCprop}, we conclude that the state $\Delta_\ell(t,i)=0$ is positive recurrent, 
thereby proving Lemma \ref{lemma:markov}.

\begin{theorem}\label{thm:DTMCprop}
	\textnormal{[Theorem 2.4-2.5 in \cite{kumar2012discrete}]} 
	If $\lim_{m\to \infty}\sum_{j=1}^{m}f_{\gamma\gamma}^{(j)}=\infty$, then the state $\gamma$ is recurrent. Additionally, if $\lim_{m\to\infty}\frac{1}{m}\sum_{j=1}^{m}f_{00}^{(j)}>0$, then $\gamma$ is positive recurrent.
\end{theorem}

\section{Proof of Theorem \ref{thm:utilfn}} \label{appendix:proof-utilfn}

If the learning algorithm converges to the maximizer $p_\ell^*$, then it should satisfy: 
\begin{gather} \label{eq:pl*inExpAlgo1}
p_\ell^*=max\left\{p_\ell^{min}, p_\ell^*+\kappa(t)\left(e^{-\alpha_{\ell} p_\ell^* }-\frac{1}{b_\ell/p_\ell^*}-p_\ell^*\right) \right\}, \text{and} \\ \label{eq:pl*inExpAlgo2}
\frac{\partial U_\ell(p_\ell(t);\mathbf{P}_{-\ell}(t))}{\partial p_\ell(t)}\Bigg|_{p_\ell(t)=p_\ell^*}=0.
\end{gather}
Therefore using \eqref{eq:pl*inExpAlgo1} and \eqref{eq:pl*inExpAlgo2}, we can write 
\begin{equation}\label{eq:delU1}
\frac{\partial U_\ell(p_\ell(t);\mathbf{P}_{-\ell}(t))}{\partial p_\ell(t)}= e^{-\alpha_{\ell} p_\ell(t)}-\frac{1}{b_\ell/p_\ell(t)}- p_\ell(t) .
\end{equation}
Integrating on both sides of \eqref{eq:delU1} w.r.t. $p_\ell$, we get \eqref{eq:utilfndummy} (with $(1+\alpha_\ell)/\alpha_\ell$ as integration constant), which is continuous and strictly concave  ($\because\frac{\partial^2 U_\ell}{\partial p_\ell^2}<0$) for $p_\ell$ in interval $(0,1]$. Also, $\frac{\partial U_\ell}{\partial p_\ell}$ is continuous, and it can be verified that for $p_\ell=0^+$, $\frac{\partial U_\ell}{\partial p_\ell}>0$, while for $p_\ell=1$, $\frac{\partial U_\ell}{\partial p_\ell}<0$. So, $\exists p_\ell^*\in(0,1)$ at which $\frac{\partial U_\ell}{\partial p_\ell}=0$, and $p_\ell^*$ is the unique maximizer because of strict concavity of $U_\ell$.

However, note that on solving \eqref{eq:pl*inExpAlgo2}, we get
\begin{align} \label{eq:delU1p*=0}
e^{-\alpha_{\ell} p_\ell^{*}}=p_\ell^{*}\left(1+\frac{1}{b_\ell}\right).
\end{align}
Since $b_{\ell}\ge 1$, so using \eqref{eq:delU1p*=0}, we get $e^{-\alpha_{\ell}p_{\ell}^{*}}\le 2p_{\ell}^{*}$, and as $e^{-\alpha_{\ell}}\le e^{-\alpha_{\ell}p_{\ell}^{*}}$, 
therefore, $e^{-\alpha_{\ell}}/2 \le p_\ell^{*}$. Hence, $p_\ell^*\in[e^{-\alpha_{\ell}}/2,1)$.

\begin{remark}
	Note that \eqref{eq:pl*inExpAlgo1} follows from \eqref{eq:ExpAlgo}, which uses Lemma \ref{lemma:av2exp} assuming the limit $m\to\infty$. Additionally, for the convergence of $p_{\ell}(t)$, we assume $t\to\infty$, and from Theorem \ref{thm:algo2br}, $\kappa(t)\to 0$ as $t\to\infty$. Therefore, for \eqref{eq:delU1} to hold, we initially take the limit $m\to\infty$, followed by the limit $t\to\infty$. If the order of the two limits is exchanged, then $\kappa(t)$ would converge to 0 before $p_{\ell}$ converges to $p_{\ell}^*$, and hence \eqref{eq:pl*inExpAlgo1} and \eqref{eq:pl*inExpAlgo2} will not be satisfied.
\end{remark}

\section{Proof of Theorem \ref{thm:BRS2NE}} \label{appendix:proof-thm-BRS2NE}
Best response strategy for the non-cooperative game $\mathcal{G}$ can be expressed as a function $f^{br}:[\mathbf{P}^{min},\mathbf{P}^{max}]\rightarrow [\mathbf{P}^{min},\mathbf{P}^{max}]$, where $\mathbf{P}^{min}$ and $\mathbf{P}^{max}$ are $N-$dimensional vectors. To prove Theorem \ref{thm:BRS2NE}, we use contraction mapping theorem: 

\begin{theorem} \label{thm:contraction}
	\textnormal{[Theorem 6.6.4 in \cite{tao2009analysis}]}
	In a metric space $(X,d)$, a function $f:X\rightarrow X$ is called a strict contraction, if there exists a constant $\gamma\in(0,1)$, such that $d(f(x),f(y))\le\gamma d(x,y)$, $\forall x,y\in X$. Additionally, if $X$ is non-empty and compact, then $f$ has a unique fixed point, i.e., there exists a unique $x^*\in X$ such that $x^*=f(x^*)$, and sequences of the form $x(t+1)=f(x(t))$ converges to $x^*$. 
\end{theorem}

Let  $[\mathbf{P}^{min},\mathbf{P}^{max}]\subset \mathbb{R}^N$ be the metric space with infinity norm as the distance metric.  Then for any $\mathbf{P}_1,\mathbf{P}_2\in[\mathbf{P}^{min},\mathbf{P}^{max}]$,
\begin{align}
d(f^{br}(\mathbf{P}_1),f^{br}(\mathbf{P}_2))&=||f^{br}(\mathbf{P}_2)-f^{br}(\mathbf{P}_1)||_{\infty}, \nonumber \\
&\stackrel{(a)}{\le}||\mathbf{J}||_{\infty}||\mathbf{P}_2-\mathbf{P}_1||_{\infty}, \nonumber \\
&= ||\mathbf{J}||_{\infty} d(\mathbf{P}_1,\mathbf{P}_2), 
\end{align}
where in (a), $\mathbf{J}$ is the Jacobian (whose elements are given by $\mathbf{J}_{lj}\triangleq\frac{\partial p_\ell^{br}}{\partial p_j}$), and the matrix norm is induced by the vector norm. Also, $[\mathbf{P}^{min},\mathbf{P}^{max}]$ is non-empty (since $\forall k, p_k^{min}<1$) and compact. Hence, to prove the existence of unique fixed point for $f^{br}$ using Theorem \ref{thm:contraction}, it is sufficient to show that $||\mathbf{J}||_{\infty}<1$. 


Now, using Lemma \ref{lemma:plmin-ub} (discussed below), we can write \eqref{eq:delU1p*=0} as 
\begin{align} \label{eq:delU1pbr=0}
e^{-\alpha_{\ell} p_\ell^{br}}=p_\ell^{br}\left(1+\frac{1}{b_\ell}\right).
\end{align}

\begin{lemma}\label{lemma:plmin-ub}
	If $p_{\ell}^{min}\le e^{-\alpha_{\ell}}/2$, then 
	$p_{\ell}^{br}=p_\ell^*$. 
\end{lemma}
\begin{IEEEproof}
	According to Theorem \ref{thm:utilfn}, $p_{\ell}^*$ is a unique maximizer of  $U_{\ell}$ in interval $(0,1]$, therefore if $p_{\ell}^*\in[p_{\ell}^{min},1]$, then according to \eqref{eq:bestResponse}, $p_{\ell}^{br}=p_{\ell}^{min}$. Also, 
	$p_{\ell}^*\ge e^{-\alpha_{\ell}}/2$. Hence, if $p_{\ell}^{min}\le e^{-\alpha_{\ell}}/2$, then 
	$p_{\ell}^{br}=p_\ell^*$.
\end{IEEEproof}

Differentiating \eqref{eq:delU1pbr=0} w.r.t. $p_j$ $(\forall j)$ , we get
\begin{align}
\frac{\partial p_\ell^{br}}{\partial p_\ell}=0, \hspace{1ex}\text{and} \hspace{1ex}\frac{\partial p_\ell^{br}}{\partial p_j}\Bigg|_{j\ne \ell}&=\frac{e^{-\rho_{\ell 2}}\prod_{k \ne \ell,j}(1-p_k)}{(\alpha_{\ell} +\frac{1}{p_\ell^{br}})(1+\frac{1}{b_{\ell}})}.
\end{align}
Since $||\mathbf{J}||_{\infty}=\underset{\ell}{max}\left(\sum_{j}|\mathbf{J}_{\ell j}|\right)=\underset{\ell}{max}\left(\sum_{j}|\frac{\partial p_\ell^{br}}{\partial p_j}|\right)$, hence
\begin{align}\label{eq:Jinf}
||\mathbf{J}||_{\infty}&\le 
\underset{\ell}{max}\left\{\frac{e^{-\rho_{\ell 2}}}{\alpha_{\ell}+1}\sum_{j\ne \ell}\left(\prod_{k \ne \ell,j}(1-p_k)\right)\right\}, \nonumber \\
&\le 
\underset{\ell}{max}\left\{\frac{(N-1)(1-p_{global}^{min})^{(N-2)}}{e^{\rho_{\ell 2}}(\alpha_{\ell}+1)}\right\},
\end{align}
where $p_{global}^{min}\le\underset{j}{min}\{p_{j}^{min}\}$. Hence, $||\mathbf{J}||_{\infty}<1$ if \eqref{eq:Jinf} is less than 1, thereby proving the existence of a unique fixed point. Now, note that any fixed point of $f^{br}$ is also NE of $\mathcal{G}$, and vice-versa. Therefore, there exists a unique NE. Hence, $f^{br}$ (best response strategy) converges to the unique NE, thereby proving Theorem \ref{thm:BRS2NE}.

\section{Proof of Lemma \ref{lemma:convBRS2NE}} \label{appendix:proofconvBRS2NE}

Let for each node $\ell$, $p_{\ell}^{min}=p_{global}^{min}$. To satisfy the condition in Lemma \ref{lemma:plmin-ub}, we restrict $p_{\ell}^{min}$ to the interval $(0,e^{-\alpha_{\ell}}/2]$. Therefore, $p_{global}^{min} \le e^{-\alpha_{\ell}}/2 = e^{-\rho_{\ell 1}c_{\ell}}/2$. Hence, $\rho_{\ell 1}\le -\frac{1}{c_{\ell}}\ln(2p_{global}^{min})$. Note that $\rho_{\ell 1}>0$ for $p_{global}^{min}<0.5$.


Now, given that $p_{\ell}^{min}$ and $\rho_{\ell 1}$ is fixed, consider the function $f(n)=(n-1)(1-p_{global}^{min})^{(n-2)}$, where $n\in\mathbb{R}_+$. If $f(n)/(e^{\rho_{\ell 2}}(\alpha_{\ell}+1))<1$ for every $n\in\mathbb{R}^+$, then \eqref{eq:BRS2NE} is always satisfied irrespective of $N$. 

Let $n^*$ be the maximizer of $f(n)$. Therefore, 
\begin{align}
n^*&=\underset{n}{argmax}f(n)=1-\frac{1}{\ln(1-p_{global}^{min})}, 
\hspace{1ex} \text{and} \\
f(n^*)&=\underset{n}{max}f(n)=(n^*-1)(1-p_{global}^{min})^{(n^*-2)}.
\end{align}
So, $f(n)/(e^{\rho_{\ell 2}}(\alpha_{\ell}+1))\le f(n^*)/(e^{\rho_{\ell 2}}(\alpha_{\ell}+1))$, and $f(n^*)/(e^{\rho_{\ell 2}}(\alpha_{\ell}+1))<1$ is implied by $\rho_{\ell 2}>\ln(f(n^*)/(\alpha_{\ell}+1))$. Also, $\rho_{\ell 2}>0$. Hence, \eqref{eq:BRS2NE} is always satisfied if $\rho_{\ell 2}>max\{0,\ln(f(n^*)/(\alpha_{\ell}+1))\}$. 

\section{Proof of Theorem \ref{thm:algo2br}} \label{appendix:proof-thm-algo2br}

To prove Theorem \ref{thm:algo2br}, we use Theorem \ref{thm:intermediateAlgo2br}, which is a special case of Theorem 6.2 in \cite{ermoliev1988numerical}.
\begin{theorem}\label{thm:intermediateAlgo2br}
	In the optimization problem \eqref{eq:bestResponse}, 
	let $U_\ell$ be a strictly concave, continuous one-dimensional function in $p_\ell$, and $p_\ell^*$ be the unique maximizer. The stochastic subgradient method \eqref{eq:subgradAlgo} will have $\lim_{t\to\infty}p_\ell(t)=p_\ell^*$ with probability 1, if the following conditions are satisfied:
	\begin{enumerate}
		\item $U_\ell(p_\ell^*;\mathbf{P}_{-\ell})-U_\ell(p_\ell(t);\mathbf{P}_{-\ell})\leq \mathbb{E}\{v_\ell(t)|p_\ell(1),p_\ell(2),...,p_\ell(t);\mathbf{P}_{-\ell}\}(p_\ell^*-p_\ell(t))+r_o(t)$, where $r_o(t)$ may depend upon $p_\ell(1),p_\ell(2),...,p_\ell(t)$. 
		\item $\kappa(t)>0, \forall t$ and $\sum_{t=1}^{\infty}\kappa(t)=\infty$.
		\item $\sum_{t=1}^{\infty}\mathbb{E}[\kappa(t)|r_o(t)|+\kappa^2(t)|v_\ell(t)|^2]<\infty$.
	\end{enumerate}
\end{theorem}

From Theorem \ref{thm:utilfn}, we know that $U_\ell$ is a strictly concave and continuous one-dimensional function in $p_\ell\in [p_{\ell}^{min},1]$ (for fixed $\mathbf{P}_{-\ell}$), and $p_\ell^*$ 
is its unique maximizer. Therefore, according to Theorem \ref{thm:intermediateAlgo2br}, \eqref{eq:subgradAlgo} (and hence, \eqref{eq:algo}) will converge to $p_\ell^*$ almost surely if the three conditions are satisfied. Note that, assuming $r_o(t)=0,\forall t$, and using \eqref{eq:subgrad2slope} and strict concavity of $U_\ell$, condition 1 is satisfied. 

Further, if the sequence $\{\kappa(t)\}_{t\in\mathbb{N}}$ is chosen such that $\forall t$, $\kappa(t)> 0$, $\sum_{t=1}^{\infty}\kappa(t)=\infty$, and $\sum_{t=1}^{\infty}\kappa^2(t)<\infty$, then condition 2 is satisfied. 

For $r_o(t)=0$, condition 3 simplifies to $\sum_{t=1}^{\infty}\mathbb{E}[\kappa^2(t)|v_\ell(t)|^2]<\infty$. And we have $|v_\ell(t)|< \infty$. Therefore, there exists a constant $M<\infty$ such that $|v_\ell(t)|< M$. Hence, $\sum_{t=1}^{\infty}\mathbb{E}[\kappa^2(t)|v_\ell(t)|^2] \le M^2\sum_{t=1}^{\infty}\mathbb{E}[\kappa^2(t)]<\infty$ (because $\kappa(t)$ is fixed, therefore $\mathbb{E}[\kappa^2(t)]=\kappa^2(t)$, and we chose $\{\kappa(t)\}_{t\in\mathbb{N}}$ such that $\sum_{t=1}^{\infty}\kappa^2(t)<\infty$).
Hence, condition 3 is also satisfied. 

Therefore, if $ \forall t$, $\kappa(t)> 0$, $\sum_{t=1}^{\infty}\kappa(t)=\infty$, and $\sum_{t=1}^{\infty}\kappa^2(t)<\infty$, then the learning algorithm \eqref{eq:algo} converges to $p_{\ell}^*$ almost surely. Further, due to Lemma \ref{lemma:plmin-ub}, we have $p_{\ell}^{min}=p_{\ell}^{br}$, thereby proving Theorem \ref{thm:algo2br}. 

\section{Proof of Theorem \ref{thm:PoA-unity}} \label{appendix:proof-PoA-unity}

Let $\mathbf{P}^{NE}$ be the transmission probability of nodes at NE. Using \eqref{eq:delU1pbr=0} for each node $\ell$, we get
\begin{align}\label{eq:delU1_NE}
e^{-\alpha_{\ell} p_\ell^{NE}}=p_\ell^{NE}\left(1+\frac{1}{b_\ell^{NE}}\right),
\end{align}
where $b_\ell^{NE}=\prod_{k \ne \ell}(1-p_k^{NE})^{-1}e^{\rho_{\ell 2}}$.  
Further, the overall utility of the system is given by $U_{sys}(\mathbf{P})=\sum_\ell U_{\ell}(p_{\ell};\mathbf{P}_{-\ell})$. 
Therefore for each node $j$,
\begin{align}\label{eq:delU1_sys}
\frac{\partial U_{sys}(\mathbf{P})}{\partial p_j}=e^{-\alpha_{j} p_j}-p_j\left(1+\frac{1}{b_j}\right)+\sum_{\ell\ne j}\frac{p_\ell^2}{2}e^{-\rho_{\ell 2}}\prod_{k \ne \ell,j}(1-p_k).
\end{align}
Note that from Theorem \ref{thm:algo} and Theorem \ref{thm:NE-fixedPt}, we have $\mathbf{P}_{LA}=\mathbf{P}^{NE}$ (where $\mathbf{P}_{LA}$ denotes the vector of (converged) transmission probabilities obtained using the learning algorithm \eqref{eq:algo}). Therefore, 
\begin{align}\label{eq:delU1_sys_at_PLA}
\frac{\partial U_{sys}(\mathbf{P})}{\partial p_j}\Bigg|_{\mathbf{P}=\mathbf{P}_{LA}} &= \frac{\partial U_{sys}(\mathbf{P})}{\partial p_j}\Bigg|_{\mathbf{P}=\mathbf{P}^{NE}}, \nonumber \\
&\stackrel{(a)}{=} \frac{1}{2}\sum_{\ell\ne j}(p_\ell^{NE})^2e^{-\rho_{\ell 2}}\prod_{k \ne \ell,j}(1-p_k^{NE}),
\end{align}
where, $(a)$ is obtained using \eqref{eq:delU1_NE} and \eqref{eq:delU1_sys}.  

Also, for each node $j$, $U_{sys}(\mathbf{P})$ is continuously differentiable in $p_j$, and 
\begin{align}\label{eq:delU2_sys}
\frac{\partial^2 U_{sys}(\mathbf{P})}{\partial p_j^2}=-\alpha_{j}e^{-\alpha_{j} p_j}-\left(1+\frac{1}{b_j}\right)<0. 
\end{align}
So, $U_{sys}(\mathbf{P})$ is strictly concave in $p_j$ for each node $j$, and hence for a given $\mathbf{P}_{-j}$, $U_{sys}(\mathbf{P})$ is maximum for $p_j\in[0,1]$ at which the absolute value of its slope \eqref{eq:delU1_sys} is minimum (i.e., close to 0). 
Since $U_{sys}(\mathbf{P})$ is maximum at $\mathbf{P}=\mathbf{P}_{OPT}$ (in \eqref{eq:def-PoA}, we assumed $\mathbf{P}_{OPT}$ to be the optimal transmission probability vector which maximizes $U_{sys}(\cdot)$), therefore for 
$\mathbf{P}_{LA}$ (i.e., $\mathbf{P}^{NE}$) to be close to $\mathbf{P}_{OPT}$ (and $PoA\approx 1$ according to \eqref{eq:def-PoA}), \eqref{eq:delU1_sys_at_PLA} must be close to 0 for each node $j$. However, when $N$ is small (less than $4$ in Fig. \ref{fig:PoA}), then with addition of every new node in the system, number of positive terms in the summation on RHS of \eqref{eq:delU1_sys_at_PLA} increases, thereby taking the value of \eqref{eq:delU1_sys_at_PLA} far from 0 for each node $j$. Therefore, $PoA$ increases. 
But because $(1-p_k^{NE})<1$ ($\because \forall k,$ $p_k^{NE}\in(0,1)$), therefore if $N$ increases, then for each $\ell$, $(p_\ell^{NE})^2e^{-\rho_{\ell 2}}\prod_{k \ne \ell,j}(1-p_k^{NE})$ (i.e., each term in the summation on RHS of \eqref{eq:delU1_sys_at_PLA}) decreases exponentially. Hence when $N$ is large (e.g., in Fig. \ref{fig:PoA}, for $N\ge 4$), with addition of each new node, the overall value of \eqref{eq:delU1_sys_at_PLA} decreases to a value close to 0, and as a consequence, $\mathbf{P}_{LA}$ moves closer to 
$\mathbf{P}_{OPT}$. Hence, as $N\rightarrow\infty$, $PoA$ approaches unity. 

\end{document}